\documentclass[conference]{IEEEtran}
\makeatletter
\def\ps@headings{%
	\def\@oddhead{\mbox{}\scriptsize\rightmark \hfil \thepage}%
	\def\@evenhead{\scriptsize\thepage \hfil \leftmark\mbox{}}%
	\def\@oddfoot{}%
	\def\@evenfoot{}}
\makeatother
\usepackage{cite}
\usepackage{times,cite,amsmath,amssymb,amsfonts,epsfig,algorithm,algorithmic,subfigure,bold-extra,mathrsfs,color,soul}
\usepackage{amsmath,amssymb,amsfonts}
\usepackage{algorithm}
\usepackage{algorithmic}
\usepackage{graphicx}
\usepackage{textcomp}
\usepackage{subfig}

\graphicspath{{./figures/}}

\begin{document}
\title{Blockchain: Emerging Applications and Use Cases}

\author{\IEEEauthorblockN{Danda B. Rawat,  Vijay Chaudhary  and Ronald Doku}
	\IEEEauthorblockA{Data Science and Cybersecurity Center (DSC$^2$)\\ Department of Electrical Engineering and Computer Science \\ Howard University,
		Washington, DC 20059, USA \\
		E-mail: danda.rawat@howard.edu}
}
\maketitle
\begin{abstract} 
	 
Blockchain \textit{also known as} a distributed ledger technology stores different transactions/operations in a chain of blocks in a distributed manner without needing a trusted third-party. Blockchain is proven to be immutable which helps for integrity and accountability, and, to some extent, confidentiality through a pair of public and private keys. Blockchain has been in the spotlight after successful boom of the Bitcoin. There have been efforts to leverage salient features of Blockchain for different applications and use cases. This paper present a comprehensive survey of applications and use cases of Blockchain technology. Specifically, readers of this paper can have thorough understanding of applications and user cases of Blockchain technology. \\     
\end{abstract}

\section{Brief Overview of Blockchain}
\begin{figure}[!b]
	\centering
	%	\vspace{-0.4cm}     
	\includegraphics[width=3in]{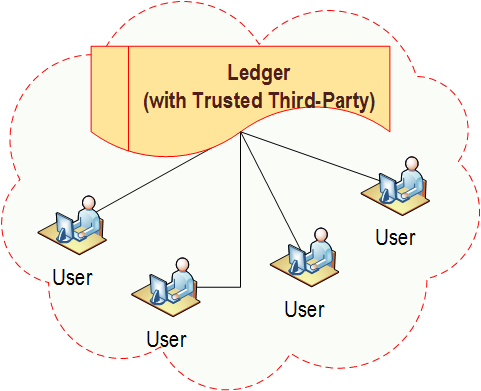} 
	\caption{Traditional centralized ledger technology with a trusted third-party.}
	\label{fig.t1}
\end{figure}

\begin{figure}[!b]
	\centering
	\includegraphics[width=3.5in]{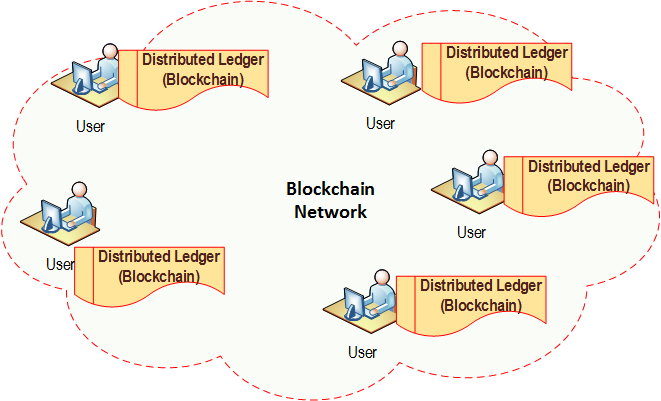} 
	\caption{A Typical Example of Blockchain Technology  -- distributed ledger technology -- without a trusted third-party.}
	\label{fig.bc}
\end{figure}

Blockchain technology is the underlying mechanism for cryptocurrencies such as Bitcoin \cite{b2}. Bitcoin, the cryptocurrency introduced in 2009, peaked a record high valuation in the December of 2017 \cite{b1} and created a hype around digital currency. Since the debut of Bitcoin, there has been several cryptocurrencies in the market holding a market cap in billions of dollars \cite{b8}.  Blockchain was first introduced in 2008 and implemented as the infrastructure of Bitcoin in 2009 by Satoshi Nakamoto, an unknown person or a group \cite{b2}. Blockchain is essentially a ``distributed ledger or database'' where all the transactions are documented regarding all the participating parties. Blockchain is a chronological chain of blocks, where each block can be considered as a page in a ledger. The chain grows continuously as miners discover new blocks and append to the existing Blockchain. Each transaction is broadcasted in the network via cryptographic communication while miners would try to collect as many transactions as they can and verify them using ``proof-of-work'' and create a new block. Miners would compete with each other to create such blocks. Once a winning block is appended to the Blockchain, a new copy of the block is broadcasted to the entire network, thus, creating a decentralized public ledger. While miners are responsible to verify transactions and update the Blockchain, they are incentivized with rewards. 
Note that the traditional ledger technologies need a trusted third-party such as bank, as shown in Fig. \ref{fig.t1}. However, the Blockchain based technology runs on peer-to-peer network, as shown in Fig. \ref{fig.bc}, where a centralized trusted third party is not needed for managing the transactions. Since, the issues such as double-spending is mitigated through consensus of miners, this system does not require intermediary, that is, a centralized trusted third party, as shown in Fig. \ref{fig.bc}. 

\begin{figure*}[!h]
	\centering
	\includegraphics[width=7in]{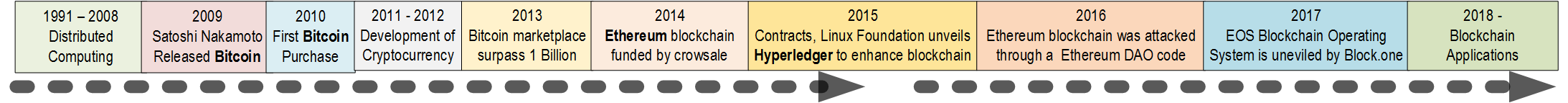} 
	\caption{The History and Milestones of Blockchain Technology.}
	\label{fig.his}
\end{figure*} 
Although Blockchain is widely used for cryptocurrencyies such as Bitcoin, this technology can also be applied to other applications. The Blockchain technology enables financial services without having financial institutions such as bank or other intermediary involved, as shown in Fig. \ref{fig.bc}. It can be implemented to conduct services such as online payment, digital assets and remittance \cite{rawat2018smart}. 
The key features of Blockchain technology -- decentralization, immutability, integrity and anonymity -- make it applicable to non-financial domains such as smart contracts \cite{b3}, the Internet-of-Things \cite{b4}, reputation systems \cite{b5}, security services \cite{b6, rawat2018ishare, malomo2018next, adebayo2019blockchain}, wireless network virtualization \cite{rawat2018leveraging} and so on.

In this paper, we briefly discuss the history and architecture of Blockchain followed by its applications and use cases in different domains. 
Although the technology has been widely praised and discussed in academia and industry for different applications, a comprehensive documentation of its emerging applications and use cases are rarely found in the literature.

\section{Brief History of Blockchain Technology}
Although the technologies involved in Blockchain such as cryptographically secured chain of blocks \cite{b9} and Merkle trees \cite{b10} were developed in the early 1990s, the first Blockchain was conceptualized and implemented Satoshi Nakamoto in 2008 \cite{b2}. The work was published on a paper entitled ``Bitcoin: A Peer-to-Peer Electronic Cash System.'' The paper introduced a peer-to-peer version of digital cash that can function with having any central authority such as bank to verify transactions. Bitcoin was the first implementation of this technology. After the publication of \cite{b2}, an open source program was published by the same author that began with the Genesis block of 50 coins.

\section{Block Architecture}
\begin{figure}[!t]
	\centering
	\includegraphics[width=3.3in]{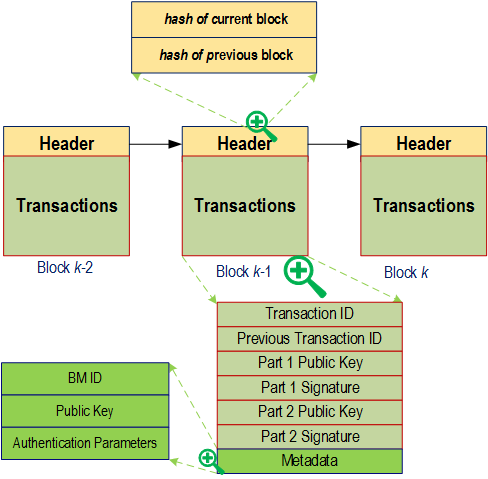} 
	\caption{Typical Blocks with Header and Transactions  in Blockchain Technology.}
	\label{fig.block}
\end{figure} 

\begin{figure*}[!ht]
	\centering
	%	\vspace{-0.4cm}     
	\includegraphics[width=7in]{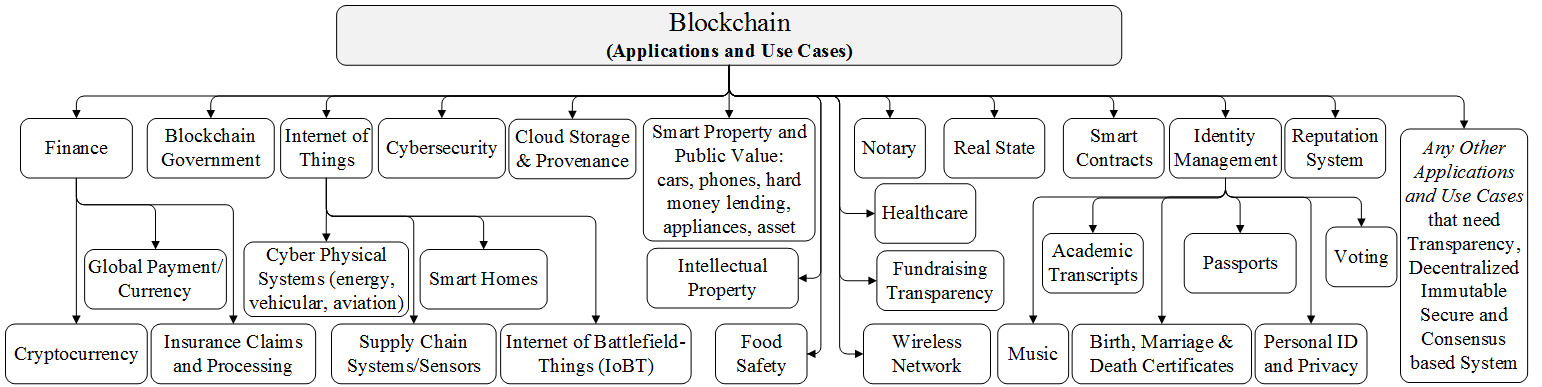} 
	\caption{Different Blockchain Applications and Use Cases.}
	\label{fig.app}
\end{figure*}

Blockchain is a chronological sequence of block where each block holds a complete list of transactions, as shown in Fig. \ref{fig.block}. It follows the data structure of linked list, where each block points to a previous block through the reference of the hash value of the previous block, also called parent block. The first block of a Blockchain, the genesis block, has no block to point to. 
A block is composed of metadata (block header) and list of transactions (block body). The metadata includes block version, parent block hash, merkle tree root hash, timestamp and nonce. Nonce is an arbitrary number which is used in the cryptographic communication within the user network. 

Digital signature is required to establish a secure communication between the users in Blockchain network. Each user is assigned with a public and a private key. When a user broadcasts a transaction to the network, the user signs the transaction using the private key. Later, the recipients use the user's public key to verify the transaction. Such cryptographic communication preserves the integrity of the transaction within the network.

 \section{Applications and Use Cases of Blockchain}
 After successful implementation of Blockchain in Bitcoin \cite{b2} because of its salient features, Blockchain has been proposed to be used in different applications and use cases, as shown in Fig. \ref{fig.his}. We present a brief overview of each domain in the following section. 
 
 \subsection{Finance}
 Conventionally, an intermediary such as a bank, verifies and processes the financial transactions. Having such a centralized system puts immense work in the hand of intermediaries, meanwhile the transactions are prone to errors as multiple uncoordinated parties are required to keep the record and adjust them. 
 Thus, the entire process is time-consuming and costly. The Blockchain simplifies such complications associated with financial services by introducing a distributed public ledger, where the transactions are verified by the miners using ``proof-of-work'' \cite{b1}. Since, each node in the Blockchain network has a copy of the updated Blockchain, there is transparency regarding the transactions, as shown in Fig. \ref{fig.bc}. Since, the blocks are chronologically arranged, once a block is added to the Blockchain with a verified transaction, the entire Blockchain is immutable. Thus, attackers cannot manipulate the transactions once it is registered into the system. In case of a conflicting Blockchain where branching might occur, miners always go for the longest chain, as the longest chain is more reliable. With such secured communication protocol and robust verification method, it creates an effective system to improve our existing financial services.

 \subsubsection{Cryptocurrency}
Cryptocurrency -- which holds a market cap in billions of dollars \cite{b8} -- has been possible with the help of Blockchain \cite{b2} which doesn't need a trusted third-party like a bank in traditional systems. All transactions are verifiable and  immutable.  
 
 \subsubsection{Global Payments (Global Currency)}
 Global payments become complicated and time-consuming because there are many intermediaries involved to verify the transactions. The entire process can be prone to error and costly. These issues arise essentially due to the centralization of the monetary transactions where institutions such as banks and other financial firms dictate processes and they are responsible to verify the transactions. The Blockchain technology reduces such complexities by introducing the decentralized public ledger and robust verification method to verify the transactions. Within this peer-to-peer network, global payments are quicker, verifiable, immutable and safer. There are several remittance companies \cite{bcC} such as Abra and Bitspark that are already using Blockchain technology for remittance services.

 \subsubsection{Insurance Claims and Processing}
 Insurance claim has been dealing with several fraudulent claims. Moreover, there must be updated policies and data associated with each claim to properly process an insurance claim which is difficult to handle in traditional approaches. With Blockchain technology, the process can be handled through Blockchain (distributed ledger technology) efficiently in a secure manner. Similarly, any fraudulent claims/transactions can be detected and dropped with a good confidence as multiple participants/miners need to agree on the validity of each transaction. This makes sure the insurers settle their claim which they deserve quickly and effectively.  
 
 \subsection{Blockchain Government}

In order to build trustworthy and effective government operations through collaborative and transparent networks, different government organizations and units can use Blockchain technology. Blockchain technology with its salient features will help provide accountability, transparency and trust among stake holders such as citizens, leaders, government officials and their different operations \cite{rawat2018smart,rawat2019cybersecurity}.
 Government is required to make its affairs transparent in order to address the accountability of its bodies. In order to do so, government might have to make a great amount of data open to the public \cite{rawat2019cybersecurity}. As per the report from McKinsey \cite{b11}, open data made available to the public in the Internet can benefit the people in an order of trillions of dollars. Several entities can use open data to expose illegitimate doings. The public can question the quality of health-care, and food supplies with given open data which eventually makes the system more fair and trust \cite{b11}. Therefore, releasing the data to the public is helpful for the economy, but it also has its own challenges to make the data public. When the data is released only once a year, it is largely left unnoticed by the public. Thus, an alternative to this can be a Blockchain government, where the data is distributed in the public ledger, and is open to the public all the time. Moreover, smart contracts can be used to ensure the electorates work in the favor of the electors. The contracts can be based on the manifesto of the electorates, and they only get paid once they meet the demand of the electors via the smart contracts. This kind of technology can put the electorates in check and possibly enforce them to fulfill their promises.

 \subsection{Internet of Things (IoT)}
 The number of electronic devices getting connected to the Internet is rapidly increasing every year \cite{b7}. With the massive number of devices interlinked to each other creates the Internet-of-Things (IoT). The IoT is expected to transform the way of lives where ideas like smart homes is feasible. While this new phenomenon is likely to make lives easier, having massive number of heterogeneous devices connected to the Internet creates graves issues regarding cyber security and  privacy. 
 The Blockchain can be an important technology to secure IoT. Having millions of devices connected to each other and communicating, it is important to ensure that the information flowing through IoT remains secured and makes the participants accountable. 
 
 \subsubsection{Energy Cyber Physical System}
Smart energy grid systems are becoming complex cyber-physical systems (CPS) where complex interactions among power generation, distribution, utility offices and users happen in a bidirectional manner \cite{rawat2015cyber}. Salient features of Blockchain technology provide a secure and verifiable environment to support interactions in energy CPS \cite{zhaoyang2018blockchain}.
 
\subsubsection{Vehicular Cyber Physical System}
Vehicular cyber physical system  is regarded as the backbone technology for intelligent transportation systems and autonomous driving \cite{rawat2016vehicular,rawat2015cyber} for improving road safety and traffic efficiency. Security and privacy in vehicular cyber physical system are always central issues since vehicles are ties to the private information of their  owner, driver or renter. Blockchain with its features such as decentralization, immutability, integrity and anonymity through a pair of public and private keys can be leveraged to build a smart and secure autonomous intelligent transportation system 
 \cite{sharma2017block}. 
 
 \subsubsection{Blockchain in Aviation Systems} 
Blockchain in aviation industry can offer robust collaborative partnerships among service and product providers to offer travel services as well as products in a distributed secure way. Smart contracts could streamline the interactions among businesses and different units within the business \cite{akmeemana2017blockchain}.
 
  \subsubsection{Supply Chain Systems/Sensors}
 Smart sensors can be helpful for the companies to gather information regarding the supply chain as they are transported around the globe. Several leading supply chain companies are reported to use smart sensors to track supplies. Therefore, the number of such sensors is expected to grow rapidly in th near  future. Having such a massive distribution of sensors, there will be enormous amount of data to be collected and analyzed. Blockchain technology can be used for disruptive transformation for efficient and secure supply  chains and network \cite{korpela2017digital}.

 \subsubsection{Smart Homes }
Blockchain in the context of smart homes with IoT devices can help to have secure and reliable operations for smart home operations \cite{dorri2017blockchain}. However, implementation of Blockchain in such resource constrained IoT systems is not straightforward because of high resource demand required for proof-of-work, limited storage capacity, low latency and low scalability  \cite{dorri2017blockchain}.

 \subsubsection{Internet of Battle-field Things (IoBT)}
 Internet-of-Battle-field Things (IoBT) is regarded as the backbone for smart defense and warfare applications where Battle-field Things such as combat equipment, unnamed areal vehicles, ground vehicles, fighters with sensors can collect intelligent information to enable informed decision real-time  in a secure and immutable manner. Note that the IoBT is so diverse in a way that it consists of different devices (combat equipment, unnamed areal vehicles, ground vehicles and fighters), platforms, networks and connectivity. This diversity several challenges for secure, privacy-aware and trustworthy battlefield operations such as communication and computing. Blockchain technology can help to have secure and reliable operations for IoBT \cite{tosh2018blockchain}. 
 
 \subsection{Cybersecurity}
  Another application of Blockchain is cybersecurity where threat information can be sharing using Blockchain among participants/organizations to combat future cyber attacks \cite{rawat2018ishare, adebayo2019blockchain}. However, Blockchain will not be able to fix everything but its features can be leveraged to harden the systems against  multitude of cyber threats.

 \subsection{Smart Property and Public Value}
 All entities/property such as house, land, automobiles, stocks, etc can be represented in the ledger technology and Blockchain can be used to keep the track of all operations and property records.  Once, the records are kept in the Blockchain, it will be shared with all the concerned or participating parties which can easily be used to establish contracts and verify them. Thus, with a decentralized ledger, any lost record can be duplicated from the network and immediately can be used to recover the loss \cite{crosby2016blockchain}.

 \subsubsection{Hard Money Lending}
 Hard money lending serves people to mitigate financial burdens in a short term. It requires the borrower to have property such as real estate as a collateral. Thus, it is important that the collateral is legit and trustworthy. Lenders can lose money if the collateral is not redeemable. Similarly, the borrower might also lose its property if the lender use fraudulent policies as part of the agreement. With Blockchain, both the property and the policies can be encoded in the ledger and distributed among the users. This will create a healthy setting where people can trade with complete strangers due to the transparency and security of the Blockchain. Smart contracts can be deployed using Blockchain for this kind of scenarios.

 \subsubsection{Cars and Phones}
 Personal devices such as phones are protected using authentication keys. Similarly, cars are only accessible to the owners using smart keys. This kind of technology is possible with cryptography, and yet, such methods can fail if the authentication key is stolen or copied or transferred. Such issues can be fixed in the Blockchain ledger where users/miners can replace and replicate lost credentials.

 \subsubsection{Smart Appliances}
  Smart appliances are essentially electronic devices aided with cyber system such that the cyber portion can communicate information regarding environment around the device and the device itself. It is essentially about the idea of a ``talking toaster'' where a toaster can give its user information relevant to its usage. A home connected with smart appliances can be considered a smart home, where the cyber physical system tries to optimize the functionalities of the smart devices, providing maximum utilities to its users. With so many devices involved as part of smart appliances, we can encode them in the Blockchain as smart property. Such practice could easily ensure  the ownership of a user over these devices. 
 
\subsubsection{Asset Management}
Asset management involved multiple parties where each party is required to keep the transactions. While keeping the same transaction in different places can make the entire process inefficient and prone to errors. To make the matter worse, asset management might also involve cross-border transactions, adding more complexity to verify the transactions. Such issues can be dealt with a distributed ledger where each party can have a copy of the entire transactions and get updated about each transaction using cryptographic communications \cite{notheisen2017trading}. This improves the efficiency and reduces the cost as there would be no intermediary to verify the transactions. 
 
 \subsection{Cloud Storage and Provenance}
 Metadata that records the history of the creation and all operations including file/data accessing activities can be kept in the Blockchain which then be shared with all stakeholders. Data provenance through Blockchain is important for applications like accountability and forensics % in both collaborative and non-collaborative processes
 \cite{liang2017provchain}.

\subsection{Intellectual Property}
Intellectual Property management system could leverage the Blockchain technology to enforce provable  intellectual property rights \cite{zeilinger2018digital} where verifiable, immutable and secure operations in Blockchain could help any disputes. 

\subsection{Food Safety}
Food safety is one of the most critical issues to be addressed since over 0.6 billion (equivalently  1 in 10 people) in the world become ill  after consuming bad food every year \cite{galvin2017ibm}. About 1,167 people die every day \cite{galvin2017ibm}. 
To prevent these issues, Blockchain technology can help to prevent counterfeiting issues for food to have visibility across food supply chain and help to  access any information such as food content, its origins, expiration, etc. in seconds. Food consumers will better control over food and information with high accuracy and transparency for food safety \cite{galvin2017ibm}.  

\subsection{Blockchain Notary}
Blockchain using distributed ledger technology with cryptography replaces trusted third parties such as a notary (trust third party in the traditional systems). Blockchain helps the entire notary process by automatically executing process in a cost-effective, transparent and secure manner \cite{nofer2017blockchain}

\subsection{Blockchain Health-care}
 Personal health records are sensitive information and needs to be dealt with high security. Such personal records can be encoded and stored using Blockchain and provide a private key which would allow only specific individuals to access the records. Similarly, the same protocol can be applied to conduct research where personal records are used via HIPAA laws to ensure confidentiality of the data. Patients records in the Blockchain can be automatically sent to the insurance providers or doctor can send the health record to concerned parties securely \cite{mettler2016blockchain}.
 
\subsection{Fundraising and Transparency}
Transparency is one of the issues to be addressed in fund-raising activities to make the process trustworthy. Blockchain as a distributed ledger technology can ensure  the transparency, security, and integrity in fund-raising activities by leveraging Blockchain features such as immutability, verifiability and security \cite{zhu2016analysis}.

\subsection{Wireless Networks and Virtualization}
 Wireless network is suffering from explosive growth of IoT and CPS applications and there have been different approach studied to enhance the network capacity and coverage  \cite{rawat2015dynamic,rawat2018payoff} 
 Blockchain can be used to sublease wireless resources such as RF slices to network service providers or third party like mobile virtual network operators (MVNOs) in a verifiable way so that quality of service of the users would be met by preventing double spending/subleasing of same   wireless resource to multiple parties in a distributed manner \cite{rawat2018leveraging, rawat2017edge}.

\subsection{Real State}
Blockchain technology as a distributed ledger database system can offer benefit for the real estate industry. Property title recording can be done using blocks with transactions in Blockchain rather than using traditional/current record keeping system \cite{spielman2016blockchain}.

\subsection{Smart Contracts}
 Smart contracts digital entity written in a Turing-complete byte language, called EVM bytecode \cite{b47}. They are essentially a set of functions where each function is a sequence of instructions. Such contracts are embedded with conditional statements which enables them to self-execute. Smart contracts can be a replacement to intermediaries which make sure that all parties are obliged by the agreed terms. Thus, with Blockchain, such regulatory bodies become redundant. 
 Smart contracts based on Blockchain ensures that the participants know the contract details and the agreement are automatically implemented once the conditions are fulfilled.  In order to make the smart contracts work, there is a group of mutually ``untrusted'' peers called miners who verify the transactions related to the contract. Each transaction broadcasted to the Blockchain network is collected by the miners and verified before they are encoded to a new block and appended to the Blockchain. Any potential conflict is resolved through the consensus protocol which is based on ``proof-of-work''. Thus, a smart contract only work if there is no bias or majority in the computational power of the network, thus ensuring the decentralization in the network. The miners are rewarded for creating new blocks under the protocol everyone miners are required to follow. Any miner’s work is discredited by other miners if he/she does not follow the protocol, thus there is an incentive for each miner to follow the rules.

 \subsection{Identity Management}
 In this section, we present brief overview of different identity management based applications and how they could benefit from Blockchain technology. 
\subsubsection{Academic Records}
Blockchain can be used to store academic records for students and universities in a decentralized ledgers \cite{sharples2016blockchain}. This academic record keeping system will be temper-proof, verifiable, immutable and secure \cite{sharples2016blockchain}.

 \subsubsection{Blockchain Music}
 In music industry, it is a huge challenge to own products via ownership rights, and benefit from royalty distribution. In order to monetize digital music products, ownership rights are required. The Blockchain and smart contracts technology can be used to create a comprehensive and accurate decentralized database of music rights. Meanwhile, the ledger can be used to provide a transparent information regarding the artist royalties and real time distributions to all the labels involved. Digital currency can be deployed to make the payments as per the terms of contracts.
 
  \subsubsection{Birth, Marriage and Death Certificates}
 The record of birth, marriage and death are important records of a citizen as they are used to confirm citizenship of citizens, and grant rights as per their status such as voting rights and work permits. While keeping such records in a conventional method can be slow and prone to error, such issues can be fixed with the public ledger such as Blockchain. The Blockchain can make such records more reliable by encrypting the records \cite{sullivan2017residency,doveylove}.
 
  \subsubsection{Passports}
 The first digital passport was launched in 2014 \cite{b12} which could help the owners to identify themselves online and offline. With this Blockchain technology, a user can take a picture and share it via cryptograpphic communication, which can be used to share the picture and verify among the users via digital signatures. In the Blockchain based passport system, passports are stored in the distributed ledger, which is confirmed/verified by the users as well as government.
 
  \subsubsection{Personal Identity and Privacy}
 We perform several transactions that are based on our personal information. For instance, we can only buy alcoholic beverages, or get into bars, and several other public places depending on our age. Similarly, there is some level of personal profiling done via companies we interact with through online shopping, or personalized web surfing. Thus, there is a notion of personal identity that is being traded in the market to the advertiser so that they can target essential products to users as per their need. While such personal identity is fairly being traded in the market, it is essential to protect the privacy of the users. Therefore, the Blockchain can be used to protect the identity of the users by encrypting the data and securing it from attackers \cite{jacobovitz2016blockchain,rawat2018ishare,rawat2018smart}.
 
\textit{ Personal ID}: There are several personal identifications we carry around such as our driver's license, student identity cards, keys, social security number card, state identification card, etc. Blockchain can be used to store these identifications as digital form of IDs that will replace all forms of traditional physical identifications. Essentially, one Blockchain ID could be used for all kind of identifications used identify the same subject or object\cite{andrade2016systems}. 
 
 \subsubsection{Voting}
 Blockchain could offer many tangible benefits for verifiable secure voting system in coming years. Current voting system has flaws and hard to verify votes and votes. Thus, Blockchain with its features could provide immutable, verifiable and secure voting system where voter can cast their votes with highest confidence from anywhere in the world \cite{ernest2017blockchain,osgood2016future}.

 \subsection{Reputation System}
 Reputation system is an important measure on how much a community trusts a person. Such a system plays an important role to assess a person through their reputation, which is evaluated on his/her past transactions and interactions with the community. Credit system can be thought of as a reputation system, where users are given credit scores based on their financial activities and later they are used to make decision regarding other financial transactions. There can be falsification of such system if the integrity of the data is compromised. Thus, it is important to securely keep the record of past transaction and fairly evaluate the reputation of a users. Here, Blockchain can be a really important technology as it keeps a distributed public ledger which is scrutinized by the consensus of users in the network.

\subsection{Other Applications and Use Cases}
Blockchain technology can be used in any scenarios when a trusted third-party is not needed or peer-to-peer system is needed for managing the transactions, as shown in Fig. \ref{fig.bc},  with features like transparency, decentralization, integrity, immutability, security and privacy.  However, Blockchain has some limitations such as high delay introduced by consensus process, large size of the blocks in Blockchain, etc.    
 
\section{Summary}
This paper has briefly summarized not only how Blockchain works but also its  different emerging applications and use cases. By reading this paper, readers can have better understanding of what is a Blockchain and what are its different applications and user cases.   
 
\section*{Acknowledgment}
%{\footnotesize 
	{This work is partly supported by the U.S. Air Force Research Lab (AFRL), U.S. National Science Foundation (NSF) under grants CNS 1650831 and HRD 1828811, and by the U.S. Department of Homeland Security (DHS) under grant award number, 2017‐ST‐062‐000003. However, any opinion, finding, and conclusions or recommendations expressed in this document are those of the authors and should not be interpreted as necessarily representing the official policies, either expressed or implied, of the funding agencies.
	}
%\vspace{-0.35cm}
%\bibliography{Refs1}
%\bibliographystyle{ieeetr}

\end{document}